\documentclass[conference]{IEEEtran}

\IEEEoverridecommandlockouts

\usepackage{amsmath,amssymb,epsf,latexsym,graphicx,color}
\usepackage{mathrsfs}

\usepackage{cite}

\newcounter{lemma} 

\newcounter{corollary} 

\newcounter{proposition} 

\newcounter{remarkjs} 

\def\ccalE{{\ensuremath{\mathcal E}}}

\def\ccalH{{\ensuremath{\mathcal H}}}
\def\ccalI{{\ensuremath{\mathcal I}}}

\def\ccalN{{\ensuremath{\mathcal N}}}

\def\ccalT{{\ensuremath{\mathcal T}}}

\def\ccal0{{\ensuremath{\mathcal 0}}}
\def\bbF{{\ensuremath{\mathbf F}}}

\def\bbH{{\ensuremath{\mathbf H}}}
\def\bbI{{\ensuremath{\mathbf I}}}

\def\bbV{{\ensuremath{\mathbf V}}}

\def\bbT{{\ensuremath{\mathbf T}}}

\def\bbY{{\ensuremath{\mathbf Y}}}

\def\bbe{{\ensuremath{\mathbf e}}}

\def\bbh{{\ensuremath{\mathbf h}}}
\def\bbi{{\ensuremath{\mathbf i}}}

\def\bbq{{\ensuremath{\mathbf q}}}

\def\bbv{{\ensuremath{\mathbf v}}}

\def\bbz{{\ensuremath{\mathbf z}}}
\def\bb0{{\ensuremath{\mathbf 0}}}
\def\bbeta{{\mbox{\boldmath $\eta$}}}

\def\bbxi{{\mbox{\boldmath $\xi$}}}

\makeatletter
\def\wideubar{\underaccent{{\cc@style\underline{\mskip8mu}}}}
\def\Wideubar{\underaccent{{\cc@style\underline{\mskip6mu}}}}
\makeatother

\makeatletter
\def\widebar{\accentset{{\cc@style\underline{\mskip8mu}}}}
\def\Widebar{\accentset{{\cc@style\underline{\mskip6mu}}}}
\makeatother

\usepackage{subfigure}

\usepackage{algorithm}

\begin{document}

\title{Moving-Horizon Dynamic Power System State Estimation Using Semidefinite Relaxation}

\author{\IEEEauthorblockN{Gang Wang$^{1,2}$, Seung-Jun Kim$^{2}$, and Georgios B. Giannakis$^{2}$}
\IEEEauthorblockA{$^1$ School of Automation, Beijing Institute of Technology\\
Beijing 100081, China\\
$^2$ Dept. of ECE and Digital Tech. Center, Univ. of Minnesota\\
Minneapolis, MN 55455, USA\\
E-mail: \{wang4937,seungjun,georgios\}@umn.edu}
 \thanks{This work was supported by NSF grant 1202135, and by the Institute of Renewable Energy and the Environment (IREE) at the University of Minnesota, under grant No.~RL-0010-13. G. Wang was supported in part by the China Scholarship Council.}}

\maketitle

\begin{abstract}
Accurate power system state estimation (PSSE) is an essential prerequisite for reliable operation of power systems. Different from static PSSE, dynamic PSSE can exploit past measurements based on a dynamical state evolution model, offering improved accuracy and state predictability. A key challenge is the nonlinear measurement model, which is often tackled using linearization, despite divergence and local optimality issues. In this work, a moving-horizon estimation (MHE) strategy is advocated, where model nonlinearity can be accurately captured with strong performance guarantees. To mitigate local optimality, a semidefinite relaxation approach is adopted, which often provides solutions close to the global optimum. Numerical tests show that the proposed method can markedly improve upon an extended Kalman filter (EKF)-based alternative.

\end{abstract}


\section{Introduction}
\label{sec1}
The electric power system is a large-scale cyber-physical system, composed of thousands of physical and computational modules, spanning over a wide geographical area. The energy management system (EMS)/supervisory control and data acquisition (SCADA) systems are responsible for monitoring, control and optimization of the power grid, performing a slew of tasks including bad data detection and analysis, economic dispatch, and optimal power flow~\cite{book:abur2004, ieee:monticelli2000, spm:giannakis2013}. Accurate power system state estimation (PSSE) is an essential prerequisite for these functions, providing the operator with basic visibility to real-time states of power systems. PSSE is also critical for security assessment necessary to detect instabilities and contingencies, and to determine necessary corrective actions~\cite{spm:huang2012}.


The state of a power system refers to the complex voltages consisting of voltage magnitudes and phase angles, at all buses in the grid. Given the network topology and impedance parameters, all nodal and line electrical quantities of interest are completely characterized by the system states. The goal of PSSE is to estimate the system states from the measurements of related quantities, such as power injections and flows, and voltage magnitudes and angles, at a subset of buses. Depending on whether system dynamics are taken into account, PSSE can be divided into two paradigms: static PSSE and dynamic PSSE (also called forecasting-aided PSSE)~\cite{spm:huang2012}.

When SCADA measurements are involved, the PSSE problem becomes nonlinear and nonconvex. Traditionally, PSSE has been solved via weighted nonlinear least-squares, invoking Gauss-Newton iterations. Thus, the method is potentially susceptible to locally optimal solutions, sensitive to initialization, and troubled with convergence issues. This may become increasingly problematic in the challenging scenarios of future power systems, where system states may change significantly between measurements due to, e.g., massive integration renewables, the presence of bad data, or, cyber-attacks.

A recent progress made for mitigating these issues in the context of static PSSE is based on a semidefinite relaxation (SDR) approach, which was demonstrated empirically to yield solutions close to the globally optimal ones at polynomial-time complexity~\cite{naps:zhu2011}. SDR is well motivated in various applications in signal processing and communication~\cite{spm:luo2010}, as well as in optimal power flow problems~\cite{bai2008, tps:low2012, tsm:dall2013}. In a nutshell, the measurement model, which is nonlinear in system states $\bbv$, is re-expressed as {\em linear} in the rank-$1$ outer product $\bbV\! :=\! \bbv \bbv^\ccalH$ ($\cdot^\ccalH$ denotes Hermitian transpose), which leads to a semidefinite programming problem except for the rank-$1$ constraint. Dropping the nonconvex rank constraints yields a convex problem, from whose solutions the desired rank-$1$ solutions can be recovered using various heuristics.


While static PSSE utilizes only the measurements of current time, dynamic PSSE can leverage past measurements as well, based on the dynamical model governing the system states. The dynamics of power systems could be due to the changing frequency, or the changing line parameters. To circumvent the nonlinearity in the measurement model, approximate state estimation techniques such as the extended Kalman filter (EKF) and the unscented Kalman filter (UKF) have been advocated~\cite{HuS02, Val11}. However, such approximations can suffer from divergence due to their inability to accurately incorporate the underlying nonlinear dynamics.


Recently, moving-horizon estimation (MHE) for nonlinear dynamical systems has attracted much attention~\cite{auto:alessandri2008}, because it can provide state estimates with bounded error under appropriate assumptions~\cite{auto:alessandri2008}. Moreover, constrained MHE has been shown to offer an asymptotically stable estimator for nonlinear dynamical systems with deterministic noise terms~\cite{tac:rao2003}. Rigorous comparison of MHE and EKF for nonlinear chemical processes corroborated the robustness and improved estimation performance of the MHE method~\cite{haseltine2005critical}. When applied to the dynamic PSSE problem, however, the MHE formulation is still nonconvex, and thus difficult to yield globally optimal solutions. The key contribution of the present paper is to leverage SDR to convexify the problem and thus attain efficient near-optimal solutions.


The remainder of this paper is organized as follows. The power grid model and the (static) PSSE formulation are introduced in Section~\ref{sec2}. The SDR approach for PSSE is reviewed in Section~\ref{sec3}. In Section~\ref{sec4}, the MHE strategy for dynamic PSSE is presented and the SDR reformulation is described. The results of numerical tests are presented in Section~\ref{sec5}, and the conclusions are drawn in Section~\ref{sec6}.

\emph{Notations:} All matrices (vectors) are denoted by boldface letters. $(\cdot)^\ccalT$ and $(\cdot)^\ccalH$ represent transpose and complex-conjugate transpose, respectively; $\|\cdot\|_F$ is the matrix Frobenius norm, $\|\cdot\|$ the vector Euclidean norm, ${\rm{Tr}}(\cdot)$ the matrix trace, and ${\rm{rank}}(\cdot)$ the matrix rank; finally, $|\cdot|$ signifies the magnitude of a complex number.


\section{Modeling and Problem Formulation}
\label{sec2}
Consider a power transmission network with $N$ buses. The set of all buses is denoted by $\ccalN\!:=\!\{1,2,\ldots,N\},$ and the set of all lines by $\ccalE\!:=\!\{(m,n)\}\!\subset\!\ccalN\times\ccalN.$ In order to estimate the complex voltages $V^n$ at all buses, collected in the state vector $\bbv\!:=\!\left[V^1,V^2,\ldots,V^N\right]^\ccalT\!\!\in\!\mathbb{C}^N,$ $L$ measurements of the following types are taken: Active (reactive) power injection at bus $n,$ denoted by $P^n$ $(Q^n)$; active (reactive) power flow out of bus $m$ to bus $l,$ denoted by $P^{mn}$ $(Q^{mn})$; and voltage magnitude at bus $n,$ denoted by $|V^n|$.
Then, collect the measurements in an $L\times 1$ vector $\bbz:=\,\big[\,\{P^n\}_{n\in\ccalN_P},\,\{Q^n\}_{n\in\ccalN_Q},\, \{P^{mn}\}_{(m,\,n)\in\ccalE_P},\,\{Q^{ml}\}_{(m,\,n)\in\ccalE_Q},$ $\{|V^n|^2\}_{n\in\ccalN_V}\big]^\ccalT,$ with $\ccalN_P,$ $\ccalN_Q,$ $\ccalE_P,$ $\ccalE_Q$ and $\ccalN_V$ denoting the sets of buses or lines where the corresponding measurements are taken.

It turns out that the measured quantities $P^n, Q^n, P^{mn},$ $ Q^{mn}$, $|V^n|$ are quadratic functions of $\bbv.$ To specify this, collect injected currents at all buses in vector $\bbi:=[I^1,I^2,\ldots,I^N]^\ccalT\in\mathbb{C}^N,$ and let $\bbY\in\mathbb{C}^{N\times N}$ denote the so-called bus admittance matrix, whose entries are defined as
\begin{align}
Y^{mn}:=\left\{\begin{array}{ll}
-y^{mn},&{\text{if~}}(m,n)\in\ccalE\\
\bar{y}^{nn}+\sum_{t\in\ccalN_n}y^{nt},&{\text{if~}}m=n\\
0&{\text{otherwise}}
\end{array}\right.
\end{align}
where $y^{mn}$ is line admittance between buses $m$ and $n;$ $\bar{y}^{nn}$ the shunt admittance of bus $n$ to the ground; and $\ccalN_n$ the set of buses with transmission lines connected to bus $n.$ Upon denoting the shunt admittance at bus $n$ corresponding to line $(m,n)$ by $\bar{y}^{mn}$, the current flowing from bus $m$ to bus $n$ is given by $I^{mn}=\bar{y}^{mn}V^m+y^{mn}(V^m-V^n).$ Then the complex power injection at bus $n$ is $P^n+jQ^n=V^n(I^n)^\ccalH,$ and the complex power flowing out from bus $m$ to bus $n$ is $P^{mn}+jQ^{mn}=V^m(I^{mn})^\ccalH.$ Likewise, the squared bus voltage magnitude can also be expressed as $|V^n|^2=V^n(V^n)^\ccalH.$ Then, the measurement model is given by
\begin{align}
z^\ell=h^\ell(\bbv)+\eta^\ell,\quad \ell=1,2,\ldots, L
\end{align}
where $h^\ell(\cdot)$ is quadratic in $\bbv$, and $\eta^\ell$ is the measurement noise.

The goal of PSSE is to obtain an estimate of $\bbv$ from $\bbz.$ The static PSSE is formulated as a weighted nonlinear least-squares (LS) problem given by
\begin{align}
\hat\bbv:=&{\rm{arg}}\mathop{\rm{min}}
\limits_{\bbv}\sum_{\ell=1}^L w^\ell\big(z^\ell-h^\ell(\bbv)\big)^2\label{wnls} 
\end{align}
where $w^\ell$ represents the weight for the $\ell$-th measurement, inversely proportional to the variance of $\eta^\ell.$ Problem (\ref{wnls}) is nonlinear and nonconvex. Thus, iterative algorithms based on Gauss-Newton updates are often employed to find locally optimal solutions. Next, an SDR approach that targets globally optimal solutions is reviewed.

\section{SDR Approach for PSSE}
\label{sec3}
The idea is to re-express the quadratic function $h^\ell$ of $\bbv$ as a linear function of the  rank-$1$ matrix ${\bbV}:={\bbv\bbv^{\ccalH}}.$ Let $\bbe^n$ denote the $n$-th canonical basis of $\mathbb{R}^N,$ and define a number of admittance-related matrices
\begin{align}
\bbY^n:=&\bbe^n(\bbe^n)^{\ccalT}\bbY\nonumber\\
\bbY^{mn}:=&(\bar{y}^{mn}+y^{mn})\bbe^m(\bbe^m)^\ccalT-y^{mn}\bbe^m
(\bbe^n)^\ccalT\label{yml}
\end{align}
together with
\begin{align}
&\bbH^{P,n}:=\frac{1}{2}\left(\bbY^n\!+\!(\bbY^n)^{\ccalH}\right),\quad\!\!
\bbH^{P,mn}:=\frac{1}{2}\left(\bbY^{mn}\!+\!(\bbY^{mn})^{\ccalH}\right)\nonumber\\
&\bbH^{Q,n}:=\frac{j}{2}\left(\bbY^n\!-\!(\bbY^n)^{\ccalH}\right),\quad\!\!
\bbH^{Q,mn}:=\frac{j}{2}\left(\bbY^{mn}\!-\!(\bbY^{mn})^{\ccalH}\right)\nonumber\\
&{\text{and~}}\bbH^{V,n}:=\bbe^n(\bbe^n)^\ccalT.\label{hvn}
\end{align}
With these definitions, the following relations hold for every $n\in\ccalN$ and every $(m,n)\in\ccalE$
\begin{align}
&P^n=\,{\rm{Tr}}\big(\bbH^{P,n}\bbV\big),\qquad \; Q^n={\rm{Tr}}\big(\bbH^{P,n}\bbV\big)\nonumber\\
&P^{mn}=\,{\rm{Tr}}\big(\bbH^{P,mn}\bbV\big),\quad \; Q^{mn}={\rm{Tr}}\big(\bbH^{Q,mn}\bbV\big)\nonumber\\
{\text{and~}}&|V^n|^2=\,{\rm{Tr}}\big(\bbH^{V,n}\bbV\big)\label{vn}.
\end{align}
Then, $z^\ell$ can be expressed as
\begin{align}
z^\ell={\rm{Tr}}\big(\bbH^\ell\bbV\big)+\eta^\ell 
\end{align}
where $\bbH^\ell$ is one of $\bbH^{P,n}, \bbH^{P,mn}, \bbH^{Q,n}, \bbH^{Q,mn}$ and $\bbH^{V,n},$ corresponding to the type of the $\ell$-th measurement. Then problem (\ref{wnls}) is equivalent to
\begin{subequations}
\begin{align}
\big\{\hat\bbV\big\}:=&{\rm{arg}}\mathop{\rm{min}}
\limits_{\bbV\in\mathbb{C}^{n\times n}}\sum_{\ell=1}^{L}w^\ell\Big(z^\ell-{\rm{Tr}}\big(\bbH^\ell\bbV\big)\Big)^2\label{ssevkn}\\
{\text{s. to}}~ \bbV&\succeq \bb0,~{\text{and}}~{\rm{rank}}(\bbV)=1\label{sserank1}.
\end{align}
\end{subequations}
SDR amounts to dropping the nonconvex rank constraint in (\ref{sserank1}), yielding a convex optimization problem, which can be efficiently solved.

\section{SDR-Based MHE for Dynamic PSSE}
\label{sec4}
\subsection{MHE for Dynamic PSSE}
For dynamic PSSE, the state-space model adopted is:
\begin{subequations}
\begin{align}
\bbv_{k+1}=&\,\bbF_k{\bbv}_k+{\bbxi}_k\label{system eq}\\
\bbz_k=&\,\bbh(\bbv_k)+\bbeta_k\label{observation eq}
\end{align}
\end{subequations}
with the following notations
$$\begin{array}{ll}
k=0,1,\ldots&{\text{time~index;}}\\
{\bbv}_k\in \mathbb{C}^N&{\text{state vector~with~unknown~initial~state}}\\
&\bbv_0\in\mathscr{Y}\subseteq\mathbb{C}^N;\\
\bbF_k\in \mathbb{C}^{N\times N}&  {\text{state-transition matrix~to~be~updated~online;}}\\
{\bbxi}_k\in\mathscr{E}\subseteq\mathbb{C}^N&\text{system noise vector;}\\
\bbz_k\in\mathbb{R}^{L}&\text{measurement vector;}\\
\bbeta_k\in\mathscr{H}\subseteq\mathbb{R}^L&\text{measurement noise vector;}\\
\mathscr{Y},\mathscr{E},\mathscr{H}\!&\text{given compact sets, with }\bb0\!\in\!\mathscr{E}{\text{~and~}}\bb0\!\in\! \mathscr{H}.
\end{array}$$


Different from standard Kalman filtering set-ups, the initial state $\bbv_0,$ the process noise $\{\bbxi_k\},$ and the measurement noise $\{\bbeta_k\}$ in MHE are assumed to be unknown deterministic vectors, which take values from $\mathscr{Y},\mathscr{E},$ and $\mathscr{H},$ respectively. The constraints $\mathscr{E}$ and $\mathscr{H}$ can be interpreted as a strategy for modeling the bounded disturbances or random variables with truncated densities \cite{tac:rao2003}. We resort to the MHE strategy to perform dynamic PSSE because of well-appreciated advantages of MHE in nonlinear state estimation, such as accurate yet tractable incorporation of nonlinearities with consequent asymptotic stability. To be specific, given appropriate assumptions including that the nonlinear system is uniformly observable, and that the states belong to a compact set, MHE turns out to be an asymptotically stable observer  \cite[Prop. 3.4]{tac:rao2003}. 

The key idea behind MHE is to capitalize on a sliding window of past observations to perform state estimation. Thus, the information vector containing
$M$ past measurements as well as the current one at time $k$ is given by
\begin{align}
\mathcal{I}_k^M\triangleq \left\{\bbz_{k-M},\ldots,\bbz_k\right\}, k=M,M+1,\ldots.\label{infor vec}
\end{align}

Let $\hat{\bbv}_{k-M|k}$ denote the smoothed state estimate at time $k-M$ given $\mathcal{I}_k^M.$ The
MHE strategy focuses on obtaining $\hat{\bbv}_{k-M|k},\hat{\bbv}_{k-M+1|k},\ldots,\hat{\bbv}_{k|k}$ at any time $k=M,M+1,\ldots,$ based on the most recent
estimate $\bar{\bbv}_{k-M}:=\hat{\bbv}_{k-M|k-1}$ and $\mathcal{I}_k^M.$ This prior estimate $\hat{\bbv}_{k-M|k-1}$ is simply obtained as
\begin{align}
\hat{\bbv}_{k-M|k-1}&=\bbF_{k-M-1|k-1}\hat\bbv_{k-M-1|k-1},\nonumber\\
&\qquad\qquad\qquad k=M+1,M+2,\ldots
\end{align}
where $\bar{\bbv}_0:=\hat\bbv_{0|M-1}$ is an \emph{a priori} prediction of initial state $\bbv_0.$
Denote by $\hat\bbv_{k-M|k},\hat\bbv_{k-M+1|k},\ldots,\hat\bbv_{k|k}$ the estimates of states $\bbv_{k-M},\bbv_{k-M+1},\ldots,\bbv_{k},$ respectively, to be calculated at time $k.$ A notable simplification of the estimation scheme to obtain estimates $\hat\bbv_{k-M+1|k},\hat\bbv_{k-M+2|k},\ldots,\hat\bbv_{k|k}$ can be based upon $\hat\bbv_{k-M|k},$ through the ``noise-free`` dynamic update, that is,
\begin{align}
\hat{\bbv}_{k-M+s+1|k}=&\bbF_{k-M+s} \hat{\bbv}_{k-M+s|k}, \nonumber\\ &\qquad s=0,1,\ldots,M-1.\label{estimates}
\end{align}
Therefore, per time instant $k$, it is only necessary to determine $\hat\bbv_{k-M|k},$ since the other $M-1$ estimates can be iteratively computed via (\ref{estimates}).

Considering that the statistics of $\bbv_0,\{\bbxi_k\},\{\bbeta_k\}$ are unknown, the LS estimation criterion is given by
\begin{align}
J(\hat{\bbv}_{k\!-\!M|k};\bar{\bbv}_{k\!-\!M},\ccalI_k^M)=&\mu \big\|\hat{\bbv}_{k-M|k}-\bar{\bbv}_{k-M}\big\|^2+\nonumber\\
\sum_{s=0}^M\lambda&\big\|\bbz_{k\!-\!M+s}\!-\!\bbh(\hat{\bbv}_{k\!-\!M+s|k})\big\|^2\label{cost}
\end{align}
where the nonnegative weights $\mu$ and $\lambda$ are design parameters, tuned depending on the relative confidence in the state prediction $\bar{\bbv}_{k-M}$ and the measurements, respectively. In a nutshell, the MHE strategy can be stated as follows.

{\bf{\emph{MHE Strategy:}}}\label{problem 1} At any time $k=M,M+1,\ldots,$ given $(\ccalI_k^M,\bar{\bbv}_{k-M}),$ find estimates $\hat{\bbv}_{k-M|k},\hat{\bbv}_{k-M+1|k},\ldots,\hat{\bbv}_{k|k}$ via
\begin{equation}
\hat{\bbv}_{k-M|k}:={\rm{arg}}~\mathop{\rm{min}}\limits_{\bbv_{k-M|k}} J(\bbv_{k-M|k};\bar{\bbv}_{k-M},\ccalI_k^M)\label{problem}
\end{equation}
and (\ref{estimates}), where $\bar{\bbv}_{k-M}$ is propagated as
\begin{align}
\bar{\bbv}_{k-M}&=\bbF_{k-M-1}\hat{\bbv}_{k-M-1|k-1}, k=M\!+\!1,M\!+\!2,\ldots
\end{align}
with an initialization $\bar{\bbv}_0:=\hat\bbv_{0|M-1}.$

\subsection{SDR for MHE}

The limitation of MHE is that it requires online solutions of dynamic (nonconvex) optimization problems [cf. (\ref{problem})], which are typically solved by Gauss-Newton iterations. Instead, the fresh idea here is to leverage the SDR technique to find the near globally optimum solutions for MHE-based dynamic PSSE.

In light of (\ref{estimates}), $\hat\bbv_{k-M+1|k},\ldots,\hat\bbv_{k|k}$ can be directly calculated once we obtain the estimate $\hat\bbv_{k-M|k}.$ Similarly, by defining ${\hat\bbV_{k-M|k}}:={\hat\bbv_{k-M|k}\hat\bbv_{k-M|k}^{\ccalH}},$ we can propagate the noise-free dynamics to obtain [cf. (\ref{estimates})]
\begin{align}
\hat\bbV_{k-M+s+1|k}=&\bbF_{k-M+s}\hat\bbv_{k-M+s|k}\bbv^\ccalH_{k-M+s|k}\bbF_{k-M+s}^\ccalH\nonumber\\
=&\bbF_{k-M+s}{\hat\bbV_{k-M+s|k}}\bbF_{k-M+s}^\ccalH\nonumber\\
&\qquad\qquad\quad\quad s=0,1,\ldots,M-1.\label{matrix}
\end{align}
Since $h^\ell(\hat{\bbv}_{k-M+s|k})={\rm{Tr}}(\bbH^\ell\hat{\bbV}_{k-M+s|k})$ holds, by substituting (\ref{matrix}) into this relation, and defining $\bbT_0=\bbI, \bbT_s=\bbF_{k-M+s-1}\bbF_{k-M+s-2}\ldots\bbF_{k-M},s=1,2,\ldots,M,$ one obtains
\begin{subequations}
\begin{align}
\bbH_{s}^{P,n}:=\,&\frac{1}{2}\bbT_s^\ccalH\left(\bbY^n+(\bbY^ns)^{\ccalH}\right)\bbT_s\label{hspn}\\
\bbH_{s}^{P,mn}:=\,&\frac{1}{2}\bbT_s^\ccalH\left(\bbY^{mn}+(\bbY^{mn})^{\ccalH}\right)\bbT_s\\
\bbH_{s}^{Q,n}:=\,&\frac{j}{2}\bbT_s^\ccalH\left(\bbY^n-(\bbY^n)^{\ccalH}\right)\bbT_s\label{hsqn}\\
\bbH_{s}^{Q,mn}:=\,&\frac{j}{2}\bbT_s^\ccalH\left(\bbY^{mn}-(\bbY^{mn})^{\ccalH}\right)\bbT_s\\
{\text{and~~}}\bbH_{s}^{V,n}:=\,&\bbT_s^\ccalH(\bbe^n(\bbe^n)^\ccalT)\bbT_s\label{hsvn}.
\end{align}
\end{subequations}
Then, it can be clearly seen that
\begin{align}
h^\ell(\hat\bbv_{k-M+s|k})&={\rm{Tr}}(\bbH^\ell\hat\bbV_{k-M+s|k})={\rm{Tr}}(\bbH_s^\ell\hat\bbV_{k-M|k})\nonumber\\ &\quad s=0,1,\ldots,M,\,\, \ell=0,1\ldots,L.\label{zi}
\end{align}
Moreover, let $\bar\bbV_{k-M}:=\bar\bbv_{k-M}\bar\bbv_{k-M}^\ccalH$ be the outer product formed from the prior estimate $\bar{\bbv}_{k-M}.$ Approaches to obtain $\bar{\bbv}_{k-M}$ from $\hat{\bbV}_{k-M|k-1}$ will be discussed later.

Then, the SDR-based MHE problem can be formulated as
\begin{subequations}
\begin{align}
&\left\{\hat\bbV_{k-M|k}\right\}:={\rm{arg}}\mathop{\rm{min}}
\limits_{\bbV_{k\!-\!M\!|\!k}\in\mathbb{C}^{N\!\times\! N}}\mu \big\|\bbV_{k-M|k}-\bar\bbV_{k-M}\big\|_F^2+\nonumber\\
&\qquad\qquad\qquad\lambda\sum_{s=0}^{M}\!\sum_{\ell=1}^{L}\Big(z_{k\!-\!M+s}^\ell\!-\!{\rm{Tr}}
\big(\bbH_{s}^\ell\bbV_{k\!-\!M|k}\big)\Big)^2\label{vkn}\\
&\quad{\text{s. to}}\quad \bbV_{k-M|k}\succeq \bb0 \label{psd1}\\
&\quad\quad\quad\quad{\rm{rank}}(\bbV_{k-M|k})=1\label{rank1}
\end{align}
\end{subequations}
where
\begin{align}
&\bar\bbV_{k-M} = \bbF_{k-M-1} \left(\hat{\bbv}_{k-M-1|k-1} \hat{\bbv}_{k-M-1|k-1}^\ccalH\right)\bbF_{k-M-1}^\ccalH, \nonumber\\
& \hspace{40mm} k = M+1,M+2,...
 \end{align}
with $\bar \bbV_0 := \hat \bbv_{0|M-1} \hat \bbv_{0|M-1}^\ccalH$. The positive semidefinite constraint \eqref{psd1} together with the rank constraint \eqref{rank1} ensure that $\hat\bbv_{k-M|k}$ can be recovered from $\hat\bbV_{k-M|k}$. However, this formulation is nonconvex due to the rank constraint. Thus, SDR approach amounts to removing (\ref{rank1}) to obtain a convex optimization problem.

Due to the relaxation, optimal solution $\hat\bbV_{k-M|k}$ to the SDR problem (\ref{vkn})-(\ref{psd1}) may be of rank greater than 1. Still, one can recover  $\hat\bbv_{k-M|k}$ from $\hat\bbV_{k-M|k}$ using a number of heuristics. One way is to perform eigen-decomposition of $\hat\bbV_{k-M|k}$ as
 \begin{align}
 \hat\bbV_{k-M|k}=\sum_{i=1}^r\sigma_i\bbq_i\bbq_i^\ccalH
 \end{align}
where $r$ is the rank of $\hat\bbV_{k-M|k}$, $\sigma_1\ge\sigma_2\ge\ldots\ge\sigma_r > 0$ are the ordered eigenvalues of $\hat\bbV_{k-M|k}$, and $\bbq_i$ is the eigenvector corresponding to  eigenvalue $\sigma_i$. Then, the best rank-one approximation of $\hat\bbV_{k-M|k}$ in the LS sense is $\sigma_1\bbq_1\bbq_1^\ccalH$. Thus, $\hat \bbv_{k-M|k}$ can be approximated to $\sqrt{\sigma_1}\bbq_1$.
Another approach is to resort to randomization, where one generates random vectors according to $\ccalN(0,\hat{\bbV}_{k-M|k})$, and picks the one that minimizes the cost.
Such a randomization procedure has been empirically found to yield reasonable performance~\cite{spm:luo2010}. Once $\hat\bbv_{k-M|k}$ is obtained via any of such heuristics, estimates $\hat\bbv_{k-M+1|k},\hat\bbv_{k-M+2|k},$ $\ldots,\hat\bbv_{k|k}$ can be found again using (\ref{estimates}).

The computational complexity of solving \eqref{vkn}--\eqref{psd1} is rather high in the present form, although it is still polynomial in $N$. There are two promising directions under investigation. One is to exploit rich sparsity structure in $\bbH^\ell$ to simplify the formulation. Another direction is to consider special network structures such as the radial topology common in transmission networks. This allows second-order cone programming (SOCP) formulations, which can be solved faster~\cite{Far13}.

\section{Numerical Tests}
\label{sec5}

 \begin{figure}
\centering
\includegraphics[scale=0.62]{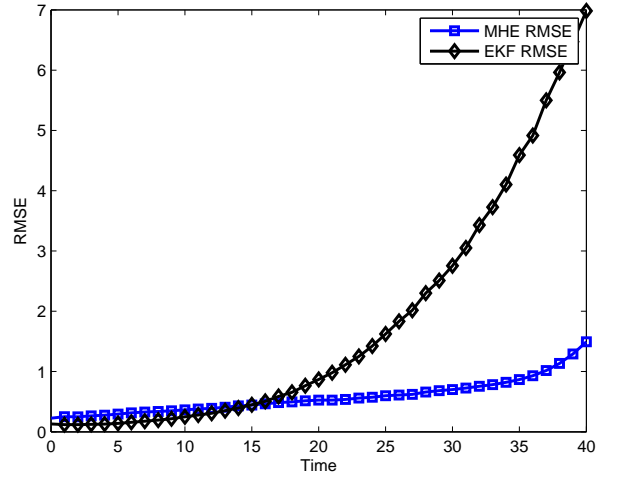}
\caption{RMSE performance comparison.}
\label{fig1}
\end{figure}

 \begin{figure}
\centering
\includegraphics[width=9.1cm,height=8.3cm]{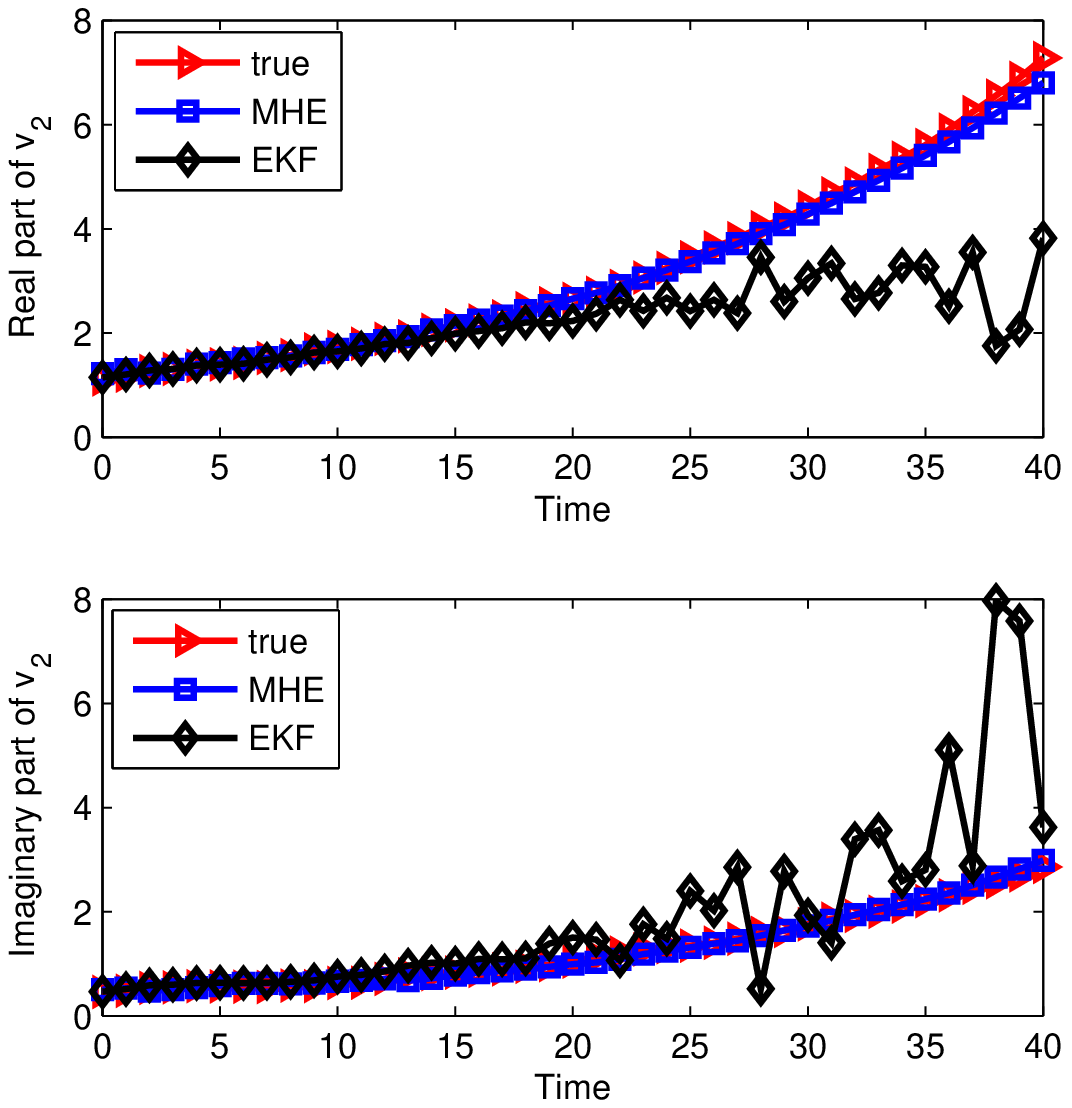}
\caption{Evolution of the real and the imaginary parts of $\bbv^{2}_k$.}
\label{fig2}
\end{figure}

 \begin{figure}
\centering
\includegraphics[width=9.1cm,height=8.3cm]{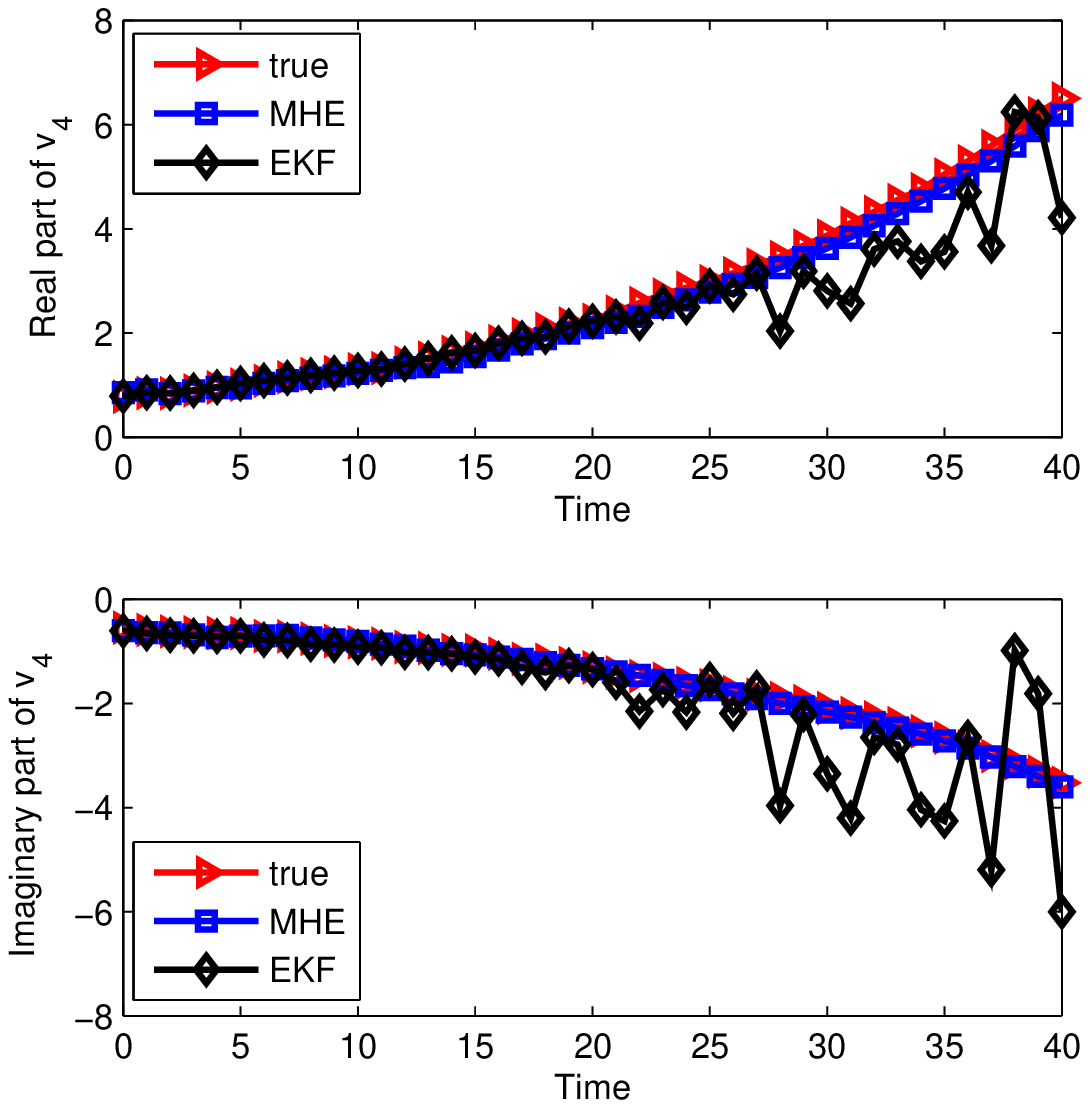}
\caption{Evolution of the real and the imaginary parts of $\bbv^{4}_k$.}
\label{fig3}
\end{figure}

 \begin{figure}
\centering
\includegraphics[width=9.1cm,height=8.3cm]{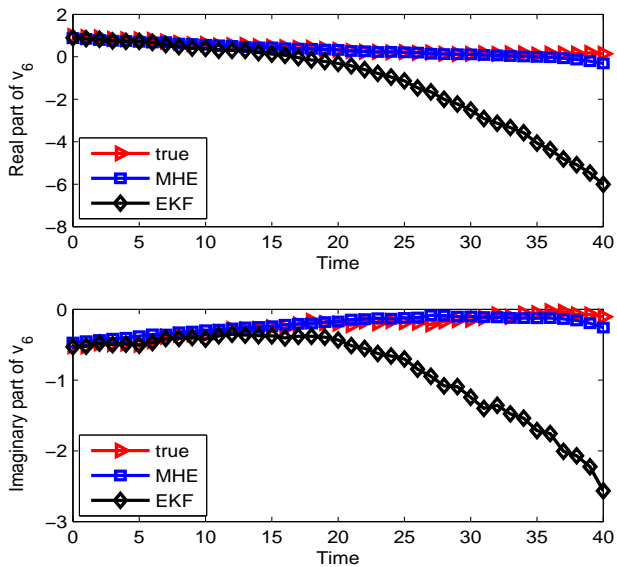}
\caption{Evolution of the real and the imaginary parts of $\bbv^{6}_k$.}
\label{fig4}
\end{figure}


The proposed SDR-based MHE approach was tested using the IEEE 6-bus system with 11 transmission lines, and compared to an existing approach that is based on the EKF \cite{HuS02}. For this, a Matlab toolbox called MATPOWER \cite{tps:thomas2011}
 was used to generate the pertinent power flows and meter measurements. To solve (\ref{vkn})-(\ref{psd1}), the CVX and SeDuMi packages were used \cite{cvx}, \cite{sedumi}.

 To simulate the slow evolution of the power system states, transition matrix $\bbF_k={\rm{diag}}(1,1.05,1.05,1.05,0.95,0.95)$ was employed and each entry of the process noise $\bbxi_k$ was generated by having both the magnitude and angle sampled according to a uniform distribution over the interval $[-0.05, 0.05].$
 The voltage magnitudes of initial state $\bbv_0$ was formed to have Gaussian distributed entries with mean $1$ and standard deviation $0.1,$ and angles uniformly distributed over $[-0.5\pi,0.5\pi].$ Bus $1$ was chosen as the reference with angle $0$ in order to fix the phase angle ambiguity \cite{naps:zhu2011}. The active and reactive power flows across lines 1-7 together with voltage magnitudes at all 6 buses were measured. Every measurement was corrupted by noise randomly generated over interval $[-0.05,0.05].$  The simulation horizon, the length of the sliding window are, respectively, $40, 2,$ and the design parameters are set to $\mu=1,$ and $\lambda=0.0075.$

Fig. \ref{fig1} compares the root-mean-square-errors (RMSEs) of the proposed approach against those of EKF, where the results were based upon 100 independent realizations. Fig. \ref{fig2}-\ref{fig4} depict the dynamic evolution of both estimates calculated by the two approaches together with the true states $\bbv_k^2$ of bus $2,$  $\bbv_k^4$ of bus $4,$ $\bbv_k^6$ of bus $6,$ respectively, where the real part is shown at the top panel and the imaginary part at the bottom. It can be clearly seen that the proposed method exhibits improved RMSE performance relative to EKF, which may even diverge from the true state depending on the initialization.

\section{Conclusion}
\label{sec6}
A dynamic PSSE algorithm has been proposed for power systems, which capitalizes on a set of recent measurements in a sliding window fashion. Since the measurement model for power grids is inherently nonlinear, traditional dynamic PSSE methods have relied on EKF/UKF approaches. Unfortunately, depending on initialization and the severity of dynamics, existing algorithms may be divergent. In contrast, the proposed approach leverages the MHE strategy and the SDR technique to accurately incorporate nonlinear dynamics, thus providing improved estimation accuracy and robustness. Numerical tests using the IEEE 6-bus system corroborated those performance claims. Further enhancements to account for false data injection as well as to reduce computational complexity are left for future work. 

\end{document}